\documentclass[a4paper,10pt,twocolumn,preprint]{elsarticle}
\usepackage[left=14mm, right=14mm]{geometry}
 \usepackage{lipsum}
 \usepackage{dblfloatfix}
 \usepackage{booktabs}
 \usepackage{mathtools}
 \usepackage{pgfplots}
\usepackage[english]{babel}
\usepackage[utf8x]{inputenc}
\usepackage{amsmath}
\usepackage{tikz}
\journal{Journal of Computational Science}

\usepgfplotslibrary{external}
\begin{document}

\twocolumn[{
\begin{frontmatter}

\author[1]{Reijers, S.A.}
\author[1]{Gelderblom, H.}
\author[2,3]{Toschi, F.}

\address[1]{Physics of Fluids Group, Faculty of Science and Technology, MESA+ Institute, University of Twente, P.O. Box 217, 7500 AE Enschede, The Netherlands}
\address[2]{Department of Applied Physics and Department of Mathematics and Computer Science, Eindhoven University of Technology, 5600 MB Eindhoven, The Netherlands}
\address[3]{IAC, CNR, Via dei Taurini 19, I-00185 Roma, Italy}

\title{Axisymmetric multiphase Lattice Boltzmann method for generic equations of state}

\begin{abstract} 
We present an axisymmetric lattice Boltzmann model based on the Kupershtokh \emph{et~al.} multiphase model that is capable of solving liquid-gas density ratios up to $10^3$. Appropriate source terms are added to the lattice Boltzmann evolution equation to fully recover the axisymmetric multiphase conservation equations. We validate the model by showing that a stationary droplet obeys the Young-Laplace law, comparing the second oscillation mode of a droplet with respect to an analytical solution and showing correct mass conservation of a propagating density wave.
\end{abstract}


\end{frontmatter}
}]
 \section{Introduction}
The lattice Boltzmann method (LBM) \cite{Benzi1992,ZhaoliGuo2014} is an efficient numerical tool to solve the Navier-Stokes equations. This numerical method can be systematically derived from the Boltzmann equations by means of a Hermite expansion approach \cite{Shan2006}. In many physically realistic flow problems one has to deal with multiphase flows such as the contact angle hysteresis of a moving droplet on a surface, a capillary rise in a cylindrical tube and droplet impact on solid surfaces. To this end, several extensions have been proposed to support multiphase flows in the LBM. In an early attempt, Gunstensen \emph{et~al.} \cite{Gunstensen1991} studied a two-component fluid lattice-gas method. Shan \emph{et~al.} \cite{Shan1993, Shan1994} were the first to incorporate intermolecular interactions to achieve phase separation in LBM. A different approach to model a multiphase fluid was developed by Swift \emph{et~al.} \cite{Swift1995}, who associated a free energy functional to the fluid. In their original form, these models lack the ability to achieve high density ratio across fluid interfaces and suffer from spurious currents near the liquid-vapor interface. In many engineering applications density ratios range from $10^1-10^3$, posing a serious limitation to the applicability of these lattice Boltzmann models in their original form. Recently, Lee \emph{et~al.} \cite{Lee2006} showed that the spurious currents are caused by discretization errors in the computation of the multiphase force. These spurious currents can be reduced to machine precision by employing a potential form of the non-ideal pressure and a isotropic central difference approximation scheme for the multiphase force. Kupershtokh \emph{et~al.} \cite{Kupershtokh2009} showed that it is possible to achieve density ratios of $10^6$ when the multiphase force is discretized by just a single-neighbour discretization scheme. The ability to achieve high density ratios makes this model applicable to many engineering applications. However, in this scheme the surface tension cannot be varied independently and spurious currents still exist. 

Recently the LBM was extended to support axisymmetric multiphase flows. These axisymmetric simulations are effectively 2D simulations in a cylindrical coordinate system. Therefore, the computational cost for axisymmetric 3D flow problems is significantly lower in comparison to the same problem in a full 3D simulation. Halliday \emph{et~al.} \cite{Halliday2001} was the first to implement an axisymmetric LBM for single-phase flows. They introduced additional source and sink terms to the evolution equation and showed that they recover the 2D axisymmetric Navier-Stokes equations. This model was improved by Lee \emph{et~al.} \cite{Lee2006Axi} who corrected a missing a source term related to the radial velocity. In addition, the method of Halliday \emph{et~al.} was extended to support non-ideal flows. Premnath \emph{et~al.} \cite{Premnath2005} were the first to implement an axisymmetric multiphase LBM. Their model is able to achieve density ratios up to $10$ and was further improved by Mukherjee \emph{et~al.} \cite{Mukherjee2007} to support density ratios up to $10^3$ and perform stable computations at lower viscosities. In this improved model, they use a pressure-evolution based LBM combined with a multiple-relaxation-time (MRT) collision model. Srivastava \emph{et~al.} \cite{Srivastava2013} developed an axisymmetric multiphase LBM based upon the widely used Shan-Chen model. In this model, they add an extra contribution to the Shan-Chen multiphase force to fully recover three-dimensionality in the system. However, large density ratios ($> 30$) could not be achieved due to the limits of the original Shan-Chen model.

In this paper, we introduce a novel and easy-to-implement axisymmetric isothermal multiphase model for high density ratio fluids. The proposed model is based on the axisymmetric LBM of Srivastava \emph{et~al.} \cite{Srivastava2013} combined with the multiphase model of Kupershtokh \emph{et~al.} \cite{Kupershtokh2009}. We show that our model can achieve density ratios up to $10^3$. Our model is discussed in detail in section II. In section III we present three validation tests. First, we verify that a stationary droplet obeys the Young-Laplace law. Then, we compare the second oscillation mode of an oscillating droplet with an incompressible analytical solution. Finally, we show that the method correctly describes the propagation of a density wave towards and away from the longitudinal z-axis. Our main conclusions and limitations of the method are discussed in section IV.
 \section{Model derivation}
 We first introduce the standard LBM. In the following sub-sections, we will gradually show the changes necessary to obtain a fully functional axisymmetric isothermal multiphase LBM. 
 \subsection{The lattice Boltzmann method}
 We use the common D2Q9 LBM, based on a two-dimensional Eulerian lattice with nine velocities. For the time evolution of the distribution function $f_i$, we use the BGK approximation with a single relaxation parameter $\tau$ \cite{ZhaoliGuo2014}. The time evolution is given by
 \begin{equation}
 \begin{multlined}
 f_i(\boldsymbol{x}+\boldsymbol{c}_i\delta t,t+\delta t) =f_i(\boldsymbol{x},t) +\\ \frac{\delta t}{\tau}(f_i^{\text{eq}}(\boldsymbol{x},t)-f_i(\boldsymbol{x},t)) + \delta t S_i(\boldsymbol{x},t),
 \end{multlined}
 \label{eq:lbm}
 \end{equation}
 where $\boldsymbol{x}$ is the position, $t$ is the time, $\delta t$ is the time step, $\tau$ is the relaxation time, $S_i(\boldsymbol{x},t)$ is a source term, $f^{eq}_i$ is the local equilibrium distribution and $c_i$ is a discrete velocity set given by
\begin{equation*}
\boldsymbol{c}_i = \begin{cases}
(0,0) &\text{$i = 0$,}\\
(1,0)_{\text{FS}} &\text{$i=(1,2,3,4)$,}\\
(\pm 1,\pm 1) &\text{$i=(5,6,7,8)$,}
\end{cases}
\end{equation*}
where the subscript $_\text{FS}$ denotes a fully symmetric set of points. The local equilibrium distribution function $f^{eq}_i$ is a second-order Taylor expansion of the Maxwell-Boltzmann distribution \cite{ZhaoliGuo2014} and is given by
\begin{equation}
 \begin{multlined}
 f^{\text{eq}}_i(\boldsymbol{x},t) = w_i\rho(\boldsymbol{x},t)\bigg[1+\frac{1}{c_s^2}(\boldsymbol{c}_i\cdot\boldsymbol{u}(\boldsymbol{x},t)) + \\ \frac{1}{2c_s^2}\left(\frac{1}{c_s^2}(\boldsymbol{c}_i \cdot \boldsymbol{u}(\boldsymbol{x},t))^2 - ||\boldsymbol{u}(\boldsymbol{x},t)||^2\right)\bigg],
 \end{multlined}
 \end{equation} 
  where $c_s^2  = \frac{1}{3}$ is the lattice speed of sound in this single-phase model and $w_i$ are the quadrature weights given by
\begin{equation*}
w_i = \begin{cases}
4/9 &\text{$i = 0$,}\\
1/9 &\text{$i=(1,2,3,4)$,}\\
1/36 &\text{$i=(5,6,7,8)$.}
\end{cases}
\end{equation*}
 The hydrodynamic quantities of the fluid, such as density $\rho$ and velocity $\boldsymbol{u}$ are calculated as weighted sums of the distribution function $f_i$
 \begin{equation}
 \rho(\boldsymbol{x},t) = \sum_i f_i(\boldsymbol{x},t),
 \end{equation}
 \begin{equation}
 \boldsymbol{u}(\boldsymbol{x},t) = \frac{\delta t\boldsymbol{F}(\boldsymbol{x},t)}{2\rho(\boldsymbol{x},t)}+\sum_i \frac{\boldsymbol{c_i}f_i(\boldsymbol{x},t)}{\rho(\boldsymbol{x},t)},
 \end{equation}
 where $\boldsymbol{u}(\boldsymbol{x},t)$ is shifted by means of an internal/external force $\boldsymbol{F}$. In the past, different implementations of a body force, $\boldsymbol{F}$, were proposed \cite{Huang2011}. Here we use the forcing scheme by Guo \emph{et~al.} \cite{Guo2002}
 \begin{equation}
 \resizebox{.9\hsize}{!}{$S_i(\boldsymbol{x},t) = w_i\left(1-\frac{\delta t}{2\tau}\right)\bigg(\frac{(\boldsymbol{c}_i-\boldsymbol{u})\cdot\boldsymbol{F}}{c_s^2}+ \frac{(\boldsymbol{c}_i-\boldsymbol{u})\cdot(\boldsymbol{c}_i-\boldsymbol{F})}{c_s^4}\bigg)$}.
 \end{equation}
 \subsection{The extension to an axisymmetric method}
 \begin{figure}[h!]
  \centering
  \includegraphics[width=0.45\textwidth]{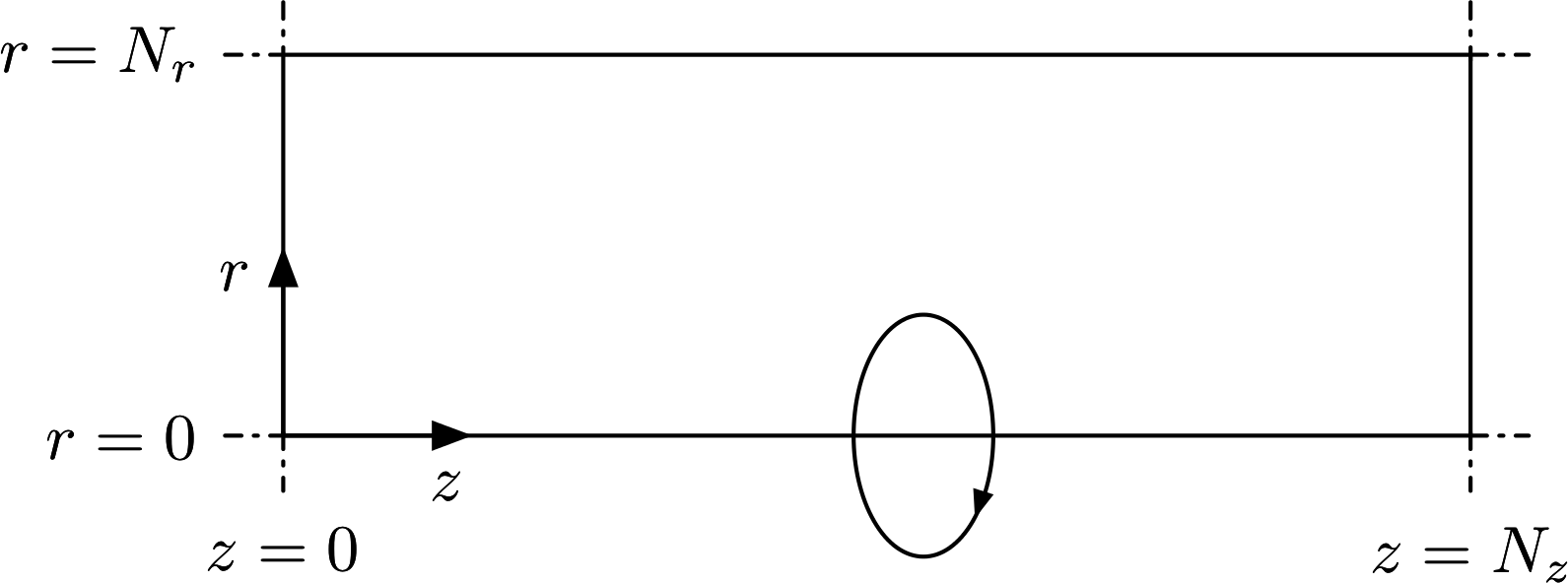}
  \caption{Schematics of the axisymmetric geometry in our LBM. The Cartesian coordinates $(x,y)$ are replaced with the axisymmetric coordinates $(z,r)$. $N_r$ is the length of the domain in the $r$-direction and $N_z$ is the length in the $z$-direction.}
\end{figure}
 In an axisymmetric flow, there is no flow in the azimuthal direction ($u_{\theta}=0$) and mass conservation reads
 \begin{align}
 \frac{\partial \rho}{\partial t} + \boldsymbol{\nabla}_{\text{c}}\cdot(\rho\boldsymbol{u}) = - \frac{\rho u_r}{r}
 \label{eq:masscylinderical}
 \end{align}
 where $\boldsymbol{\nabla}_{\text{c}} \equiv (\frac{\partial}{\partial z},\frac{\partial}{\partial r})$ is the gradient operator in a two-dimensional Cartesian coordinate system $(x\to z, y\to r)$ and $\boldsymbol{u} = (u_z, u_r)$ is the fluid velocity. The momentum equation reads
 \begin{align}
  \rho\left(\frac{\partial \boldsymbol{u}}{\partial t} + \boldsymbol{u}\cdot\boldsymbol{\nabla}_{\text{c}}\boldsymbol{u}\right) = -\boldsymbol{\nabla}_{\text{c}}P +\label{eq:momentumcylinderical} \\ \mu\boldsymbol{\nabla}_{\text{c}}\cdot\left[\boldsymbol{\nabla}_{\text{c}}\boldsymbol{u} + \boldsymbol{\nabla}_{\text{c}}\boldsymbol{u}^\text{T} \right] + \boldsymbol{C}\nonumber  
 \end{align}
 where $P$ is the fluid pressure which in a single-phase LBM is given by $P=c_s^2 \rho$ and $\boldsymbol{C}$
 \begin{align}
  C_z  = \frac{\mu}{r}\left(\frac{\partial u_z}{\partial r} + \frac{\partial u_r}{\partial z}\right), && C_r = 2\mu \frac{\partial}{\partial r}\left(\frac{u_r}{r}\right),
 \end{align}
 with $\mu$ the fluid viscosity. It is clear that equations (\ref{eq:masscylinderical},\ref{eq:momentumcylinderical}) have additional contributions to the mass and momentum conservation equations in comparison to 2D flow in the $(z,r)$-plane. These contributions ensure local conservation of mass and momentum when fluid is moving towards or away from the longitudinal $z$-axis. The single-phase LBM can be supplemented with appropriate source-terms to recover the axisymmetric conservation equations (\ref{eq:masscylinderical},\ref{eq:momentumcylinderical}) \cite{Srivastava2013}. To this end, the evolution equation (\ref{eq:lbm}) is rewritten with an additional source term $h_i$
 \begin{equation}
 \begin{multlined}
 f_i(\boldsymbol{x}+\boldsymbol{c}_i\delta t,t+\delta t) = \frac{\delta t}{\tau}(f_i^{\text{eq}}(\boldsymbol{x},t)-f_i(\boldsymbol{x},t)) + \\ f_i(\boldsymbol{x},t) +\delta t S_i(\boldsymbol{x},t)  +\delta t h_i(\boldsymbol{x}+\boldsymbol{c}_i \frac{\delta t}{2}, t+\frac{\delta t}{2}).
 \end{multlined}
 \end{equation}
 where $h_i$ is evaluated at fractional time steps. Srivastava \emph{et~al.} \cite{Srivastava2013} showed by means of a Chapman-Enskog (CE) expansion that when $h_i$ has the following form
\begin{equation}
h_i = w_i \left(-\frac{\rho u_r}{r} + \frac{1}{c_s^2}(c_{iz}H_z+c_{ir}H_r)\right),
\end{equation}
with $\boldsymbol{c}_i = (c_{iz},c_{ir})$ and 
\begin{subequations}
\begin{align}
H_z = \frac{c_{iz}}{r} \left(\mu\left(\frac{\partial u_z}{\partial r} + \frac{\partial u_r}{\partial z}\right) - \rho u_r  u_z\right),  \\
H_r = \frac{c_{ir}}{r} \left(2\mu\left(\frac{\partial u_r}{\partial r} + \frac{u_r}{r}\right) - \rho u_r^2\right),
\end{align}
\label{eq:axihz}
\end{subequations}
the resulting LBM solves the axisymmetric conservation equations given by (\ref{eq:masscylinderical},\ref{eq:momentumcylinderical}) in the limit of small Mach number. The velocity derivatives inside (\ref{eq:axihz}) are approximated by a isotropic fifth-order accurate finite difference scheme that will be given below (\ref{eq:fifthorderaccuratefinitedifference}).
 \subsection{The extension to a multiphase axisymmetric method}
\begin{figure}
	\includegraphics[width=0.45\textwidth]{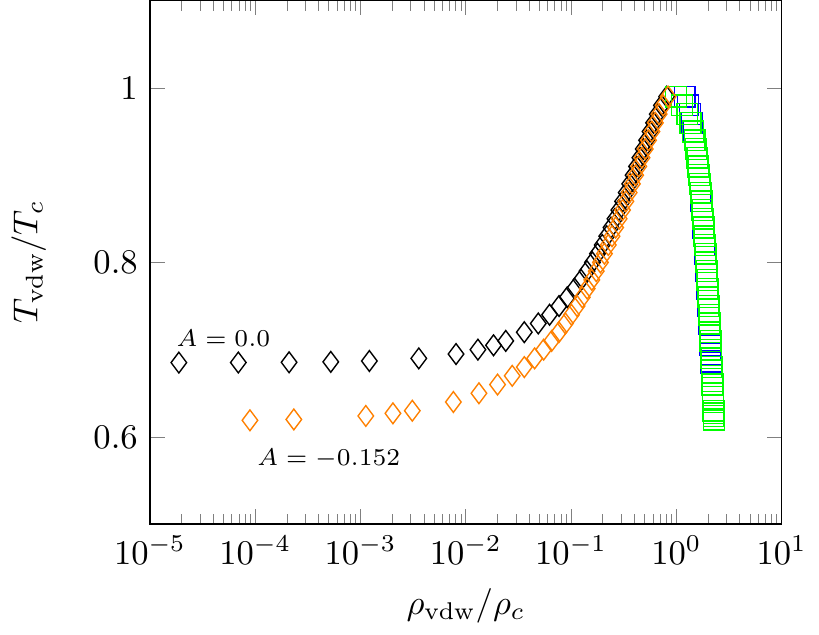}
\caption{The co-existence curve relating the equilibrium densities $\rho_\text{v}$ and $\rho_\text{l}$ to the temperature $T$ for $A=0$ (black diamonds and blue squares) and $A=-0.152$ (orange diamonds and green squares). The simulation is capable of achieving density ratios up to $10^3$ and beyond. Simulation parameters: $N_z \times N_r$ = $ 150 \times 100$, $\tau = 1.0$, $\lambda = 0.01$.}
\label{fig:coexistence}
\end{figure}
We employ a temperature-dependent body force to obtain a multiphase fluid \cite{Shan1993}. Zhang \emph{et~al.} \cite{Zhang2003} proposed a body force of the form
\begin{equation}
\boldsymbol{F}(\boldsymbol{x},t) = -\boldsymbol{\nabla}U(\boldsymbol{x},t),
\end{equation}
for which the overall effective fluid pressure in the system becomes
\begin{equation}
P(\boldsymbol{x},t) = c_s^2 \rho(\boldsymbol{x},t) + U(\boldsymbol{x},t). 
\end{equation}
In this notation, it is evident that a particular equation of state (EOS) $P_{\text{k}}$ can be obtained by simply adapting the choice for $U$ accordingly
\begin{equation}
U(\boldsymbol{x},t) =  P_{\text{k}}(\boldsymbol{x},t) - c_s^2 \rho(\boldsymbol{x},t).
\end{equation}
In this work, we will only consider the dimensionless van der Waals (vdW) EOS 
\begin{align}
P_{\text{k}} &= \lambda\left(\frac{8\rho T}{3 - \rho} - 3\rho^2\right), \label{eq:kupershtokvdwpressure}\\
c_{\text{k}} &= \sqrt{\lambda\left(\frac{24 T}{(\rho-3)^2}-6 \rho\right)} \label{eq:kupershtokhspeedofsound},
\end{align}
where $T=T_{\text{vdw}}/T_c$ is an effective temperature and $T_c$ the critical temperature, $c_{\text{k}}$ is the thermodynamic speed of sound at constant entropy, $\rho=\rho_{\text{vdw}}/\rho_c$ is the density and $\rho_c$ the critical density, $P=P_{\text{vdw}}/P_c$ is the pressure and $P_c$ the critical pressure and $\lambda = \frac{P_c}{\rho_c}\left(\frac{\delta t}{\Delta x}\right)^2$ is a scaling parameter. Kupershtokh \emph{et~al.} \cite{Kupershtokh2009} showed that for this vdW-EOS, the theoretical co-existence curve can be fully reconstructed by using the EDM forcing scheme and a special discretization for the body force
 \begin{align}
\boldsymbol{F}(\boldsymbol{x}) &= \frac{18}{3\delta t}\bigg(A\sum_{i}w_i\psi^2(\boldsymbol{x} + \boldsymbol{c}_i\delta t)\boldsymbol{c}_i + \label{eq:kupershtokhforcing} \\ &(1-2A)\psi(\boldsymbol{x})\sum_{i}w_i\psi(\boldsymbol{x} + \boldsymbol{c}_i\delta t)\boldsymbol{c}_i\bigg), \nonumber \\
 \psi(\boldsymbol{x})& = \sqrt{-P_{\text{k}} + \frac{\rho}{3}}.
\end{align}
where $A$ is a tunable parameter. For $A=0$ the scheme coincides with a local approximation scheme and for $A=0.5$ with a mean-value approximation. The vdW co-existence curve is recovered by setting $A=-0.152$ \cite{Kupershtokh2009}. 

By means of a Taylor expansion for $\psi(\boldsymbol{x} + \boldsymbol{c}_i\delta t)$ one can find a continuum expression for the body force
\begin{flalign}
\boldsymbol{F}(\boldsymbol{x}) \approx \frac{18c_s^2}{3}\bigg( \psi(\boldsymbol{x})\boldsymbol{\nabla}\psi(\boldsymbol{x}) +  \frac{(\delta t)^2c_s^2}{2}\bigg[\psi(\boldsymbol{x})\boldsymbol{\nabla}\nabla^2\psi(\boldsymbol{x}) \label{eq:taylorexpanssion} \\ + 6A\boldsymbol{\nabla}\psi(\boldsymbol{x})\nabla^2\psi(\boldsymbol{x})\bigg]\bigg) + \mathcal{O}((\delta t)^4) \nonumber.
\end{flalign}
To fully recover axisymmetry in the multiphase force, we need to use correction terms, similar to what has been done for the mass- and momentum conservation equations \cite{Srivastava2013}. In cylindrical coordinates, the Laplace operator is given by $\nabla^2 \equiv \frac{\partial^2}{\partial r^2}+\frac{\partial^2}{\partial z^2}+ \frac{1}{r}\frac{\partial}{\partial r} = \nabla^2_{\text{c}} +  \frac{1}{r}\frac{\partial}{\partial r}$. Therefore, we can rewrite (\ref{eq:taylorexpanssion}) as
\begin{flalign}
\boldsymbol{F}(\boldsymbol{x}) \approx \frac{18c_s^2}{3}\bigg( \psi(\boldsymbol{x})\boldsymbol{\nabla}_{\text{c}}\psi(\boldsymbol{x}) +  \frac{(\delta t)^2c_s^2}{2}\bigg[\psi(\boldsymbol{x})\boldsymbol{\nabla}_{\text{c}}\nabla_{\text{c}}^2\psi(\boldsymbol{x}) \\ + 6A\boldsymbol{\nabla}_{\text{c}}\psi(\boldsymbol{x})\nabla_{\text{c}}^2\psi(\boldsymbol{x})\bigg]\bigg) + \boldsymbol{F}_{\text{axis}}(\boldsymbol{x}) + \mathcal{O}((\delta t)^4) \nonumber,
\end{flalign}
where $\boldsymbol{F}_{\text{axis}}$ is given by 
\begin{flalign}
\boldsymbol{F}_{\text{axis}}(\boldsymbol{x}) = 3(\delta t)^2c_s^4\bigg(\psi(\boldsymbol{x})\boldsymbol{\nabla}_{\text{c}}\bigg[\frac{1}{r}\frac{\partial\psi(\boldsymbol{x})}{\partial r}\bigg] +\nonumber \\ \frac{6A}{r}\frac{\partial\psi(\boldsymbol{x})}{\partial r}\boldsymbol{\nabla}_{\text{c}}\psi(\boldsymbol{x})\bigg),
\label{eq:multiphasecorrections}
\end{flalign}
and can be identified as the cylindrical contribution to the multiphase force. The meaning of $c_s$ in the Taylor expansion that leads to (\ref{eq:multiphasecorrections}) is ambiguous. By imposing an EOS that differs from an ideal gas (\ref{eq:kupershtokvdwpressure}), the speed of sound becomes function of the local density and is no longer constant.  This ambiguity is a consequence of the current mathematical formulation of the LBM. We use a lattice Boltzmann scheme that is derived from a single-phase ideal gas in which the speed of sound is constant. However by imposing a non-ideal EOS, the speed of sound changes with density which invalidates its original mathematical definition. In the limit of small fluid velocity gradients, the lattice speed of sound $c_s$ in (\ref{eq:taylorexpanssion} - \ref{eq:multiphasecorrections}) may be substituted by the thermodynamic speed of sound $c_{\text{k}}$ (\ref{eq:kupershtokhspeedofsound}). As a result, we observe better consistency between the axisymmetric simulations and fully 2D simulations. Unfortunately, a mathematical proof of the validity of this substitution cannot be constructed at this moment. We leave this to future work.

The evaluation of $\boldsymbol{F}_{\text{axis}}$ requires an approximation for the derivatives of $\psi$ accurate up to order $(\delta t)^4$. Therefore, we use the following isotropic fifth-order accurate finite difference approximations  \cite{Srivastava2013}
\begin{subequations}
\begin{align}
\frac{\partial \psi(\boldsymbol{x})}{\partial r}  &= \frac{1}{36}\sum_{i=1}^8\bigg(8\psi(\boldsymbol{x}+\boldsymbol{c}_i\delta t) -  \psi(\boldsymbol{x}+2\boldsymbol{c}_i\delta t)\bigg)c_{ir} \\&+\mathcal{O}((\delta t)^5) \nonumber,\\
\frac{\partial^2 \psi(\boldsymbol{x})}{\partial r^2}  &= \frac{1}{36}\sum_{i=1}^8\bigg(\frac{8\partial\psi(\boldsymbol{x}+\boldsymbol{c}_i\delta t)}{\partial r} -  \frac{\partial\psi(\boldsymbol{x}+2\boldsymbol{c}_i\delta t)}{\partial r}\bigg)c_{ir} \nonumber \\&+\mathcal{O}((\delta t)^5) ,\\
\frac{\partial \psi(\boldsymbol{x})}{\partial z}  &= \frac{1}{36}\sum_{i=1}^8\bigg(8\psi(\boldsymbol{x}+\boldsymbol{c}_i\delta t) -  \psi(\boldsymbol{x}+2\boldsymbol{c}_i\delta t)\bigg)c_{iz} \\&+\mathcal{O}((\delta t)^5) \nonumber,\\
\frac{\partial^2 \psi(\boldsymbol{x})}{\partial r\partial z}  &= \frac{1}{36}\sum_{i=1}^8\bigg(\frac{8\partial\psi(\boldsymbol{x}+\boldsymbol{c}_i\delta t)}{\partial z} -  \frac{\partial\psi(\boldsymbol{x}+2\boldsymbol{c}_i\delta t)}{\partial z}\bigg)c_{ir} \nonumber \\&+\mathcal{O}((\delta t)^5).
\end{align}
\label{eq:fifthorderaccuratefinitedifference}
\end{subequations}
 \section{Results and Validation}
Our model is capable of achieving density ratios up to $10^3$. The co-existence curve relating the equilibrium vapor density $\rho_\text{v}$ and liquid $\rho_\text{l}$ density to the temperature $T$ is plotted in figure \ref{fig:coexistence}. In the remainder of this section we validate our axisymmetric isothermal multiphase LBM against three test cases. First, we compare the pressure difference across the liquid-vapor interface of a stationary droplet of radius $R_0$ for different temperatures $T$ to the Young-Laplace law. Then, we compare the second oscillation mode of an oscillating droplet with its analytical solution. Finally, we show that mass is correctly conserved by simulating a propagating density wave traveling towards and away from the longitudinal $z$-axis.
 \subsection{Stationary droplet validation}
 \begin{table}
  \centering
  \begin{tabular}{rcccccc}
    \toprule
    \multicolumn{1}{c}{} & \multicolumn{2}{c}{$A=0$}       & \multicolumn{2}{c}{$A=-0.152$} \\
\cmidrule(r){2-5}
\cmidrule(r){6-7}
    \multicolumn{1}{l}{$T$}  & $\sigma_{\text{lv}}^{\text{2D}}$ & $\sigma_{\text{lv}}^{\text{axis}}$ & $\sigma_{\text{lv}}^{\text{2D}}$ & $\sigma_{\text{lv}}^{\text{axis}}$ \\
    \midrule
    \multicolumn{1}{l}{0.9} & 0.305 & 0.304 & 0.371   & 0.383   \\
    \multicolumn{1}{l}{0.8} & 0.928 & 0.928 & 1.146  & 1.158  \\
    \multicolumn{1}{l}{0.7} & 1.983 & 2.009 & 2.285   & 2.318 \\
    \bottomrule
    \end{tabular}
    \caption{Surface tension evaluated using (\ref{eq:laplacepressure}). Here, $A$ denotes the tuning parameter for the multi-phase force, while $\sigma_{\text{lv}}^{\text{2D}}$ and $\sigma_{\text{lv}}^{\text{axis}}$ denote the surface tensions obtained from full (2D) and axisymmetric simulations. Simulation parameters: $N_z \times N_r$ = $ 500 \times 200$, $\tau=1.5$, $\lambda = 0.01$.}
    \label{table:laplacevalidation}
\end{table} 
 The pressure difference across the liquid-gas interface of a stationary 2D droplet with radius $R_0$ is given by the Young-Laplace equation
 \begin{equation}
 \Delta P_{\text{k}} = \frac{\sigma_{\text{lv}}}{R_0},
 \label{eq:laplacepressure}
 \end{equation}
 where $\sigma_{\text{lv}} = \sigma_{vdw}/(P_c \Delta x)$ is the surface tension between the liquid-vapor phase, $R_0 = R/\Delta x$ is the stationary droplet radius and $\Delta x = 1$ is the lattice spacing in our LBM. Note that there is only one radius of curvature in a two-dimensional system. In this validation test, we check if the value of the surface tension in the present model $\sigma_{\text{lv}}^{\text{axis}}$ is consistent with a full 2D simulation $\sigma_{\text{lv}}^{\text{2D}}$. To this end, we perform simulations of a stationary droplet with radius $R_0=150$ (lattice units) for three different temperatures $T$ and two coupling constants $A$. As boundary conditions for the axisymmetric simulation we use mid-grid specular reflection boundary condition on the longitudinal $z$-axis, periodic boundary conditions in the $r$ direction and a free-slip boundary condition at $r=N_r$. In the 2D lattice Boltzmann simulation, we used periodic boundary conditions on all sides. 
 
The results of these simulations are summarized in Table \ref{table:laplacevalidation}. When comparing the surface tension values between the full 2D and the axisymmetric simulation, we find that the maximum error is $1.3$\% for $A=0$ and $3.2$\% for $A=-0.152$, respectively. Our axisymmetric multiphase LBM is in excellent agreement with the full 2D counterpart.
 \subsection{Oscillating droplet validation}
 \begin{figure}[t]
	\includegraphics[width=0.45\textwidth]{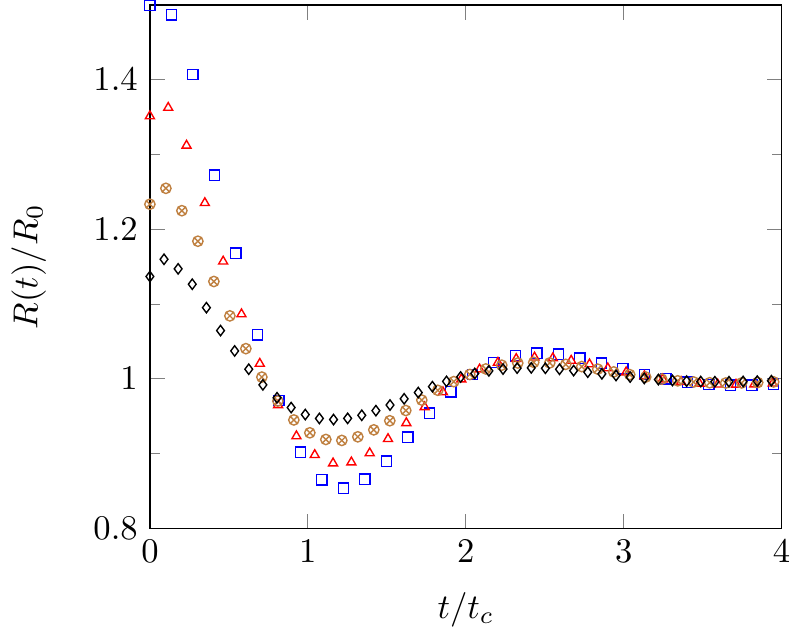}
\caption{The temporal evolution of the interfacial position along the longitudinal $z$-axis for four different initial conditions. We normalized the time by the capillary time $t_c= \sqrt{\frac{\rho R_0^3}{\lambda\sigma_\text{lv}}}$ and the droplet radius $R$ by its equilibrium radius $R_0$. Simulation parameters: $N_z \times N_r$ = $ 500 \times 250$, $\tau = 0.7$, $T=0.8$, $A=0.0$, $\lambda = 0.01$, $\rho_l/\rho_v \approx 10$.}
\label{fig:oscillatingdroplets}
\end{figure}
In this validation test we consider the problem of an oscillating axisymmetric droplet immersed in a gas. An analytical solution to the frequency and rate of damping of an oscillating droplet for arbitrary droplet radii, viscosity and surface tension was obtained by Miller and Scriven \cite{Miller1968} in the limit of an isothermal, incompressible and Newtonian fluid. Here, we consider only the second mode which is axisymmetric
 \begin{equation}
 \omega_2 =\omega_2^*-\frac{\alpha(\omega_2^*)^{1/2}}{2} + \frac{\alpha^2}{4},
 \label{eq:oscillatingfreq}
 \end{equation}
 where
  \begin{equation}
\omega_2^* =\sqrt{\frac{24\lambda \sigma_{\text{lv}}}{R_0^3(2\rho_\text{v} +3\rho_\text{l})}},
 \end{equation}
 and $\alpha$ 
   \begin{equation}
\alpha =\frac{25\sqrt{\nu_\text{l}\nu_\text{v}}\rho_\text{l}\rho_\text{v}}{\sqrt{2}R_0(2\rho_\text{v} +3\rho_\text{l})(\sqrt{\nu_\text{l}}\rho_\text{l}+\sqrt{\nu_\text{v}}\rho_\text{v})},
 \end{equation}
 with $\nu_\text{v}$ and $\nu_\text{l}$ the kinematic viscosities of the vapor and liquid phase, respectively.

 In this test, we initialized the fluid domain with an axisymmetric ellipsoidal drop, $(r/R_a)^2+ (z/R_z)^2 = 1$, where $R_a$, $R_b$ are the semi-principal lengths of the ellipsoid. The liquid and gas densities are initialized as the equilibrium densities corresponding to the temperature $T=0.8$. We measure the time dependent radius $R(t)$ of the droplet along the longitudinal $z$-axis, see figure \ref{fig:oscillatingdroplets}, for 4 different initial ellipsoid radii. Our simulation domain consists of a free-slip boundary condition at the top, periodic boundary conditions along the sides and the mid-grid specular reflection boundary condition on the longitudinal $z$-axis. The oscillation frequency $\omega_2$ can be calculated by fitting the temporal evolution of the interfacial position with that of a damped harmonic oscillator $f(t) = a + b \exp(-c t)\sin(\omega_2 t + d)$, where the interfacial position is picked as the point in space where $\rho = \frac{\rho_v + \rho_l}{2}$. We found that the frequency $\omega_2$ for the four different droplet radii with $\tau=0.7$ have a minimal relative error of $6$\% and a maximum relative error of $8$\% at density ratio $\rho_l/\rho_v \approx 10$ with respect to the analytical solution (\ref{eq:oscillatingfreq}). Relative frequency errors in oscillating droplets of the order of $8$\% are also observed in other multiphase lattice Boltzmann models \cite{Mukherjee2007,Huang2015,Premnath2008,Xu2015}. 
 \subsection{Wave propagation validation}
 \begin{figure}[h]
	\includegraphics[width=0.45\textwidth]{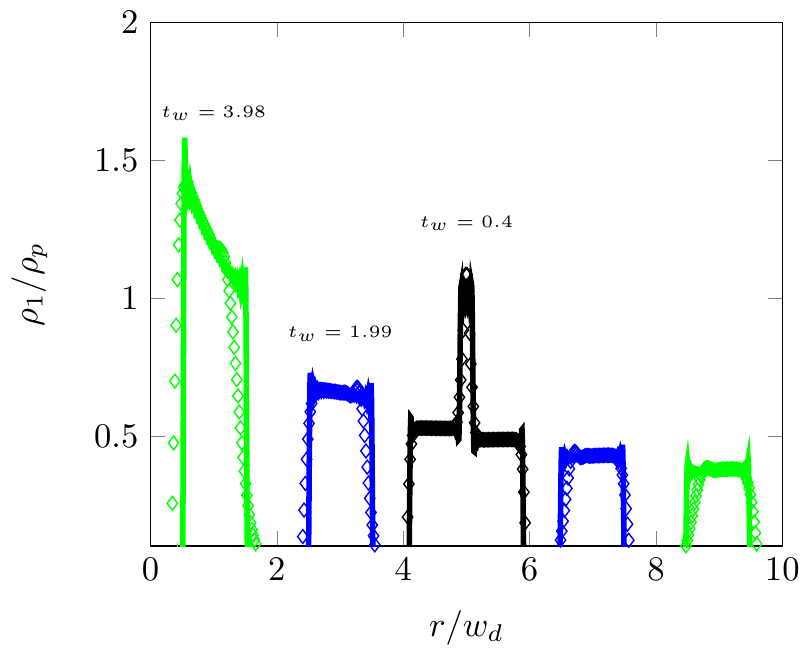}
\caption{The disturbance density field $\rho_1$ as function of the position $r$ is plotted at different time stages, where $t_w=\frac{c_k t}{w_d}$ is the dimensionless time. Simulation parameters: $N_z \times N_r$ = $ 1 \times 2000$, $\tau = 0.6$, $T=0.8$, $\rho = 1.8$, $A=0.0$, $\lambda = 0.01$, $\rho_l/\rho_v \approx 10$.}
\label{fig:propagatingdensitywave}
\end{figure}
 In the limit of small density fluctuations $\rho_1 / \rho_0 \ll 1$, the NSE can be linearized up to first order, where $\rho_1$ is a density disturbance field and $\rho_0$ the background density \cite{Skudrzyk1971}. The resulting cylindrical wave-equation for the disturbance field $\rho_1(\boldsymbol{x},t)$ in the inviscid limit reads
 \begin{equation}
 \frac{\partial^2 \rho_1}{\partial t^2} - c_k^2 \nabla_c^2 \rho_1 = \frac{c_k^2}{r}\frac{\partial \rho_1}{\partial r}.
 \label{eq:waveequation}
 \end{equation}
Here, we check if the mass of a propagating density wave towards and away from the axis is conserved and obeys the wave-equation (\ref{eq:waveequation}). The spatial temporal solution for the cylindrical wave-equation (\ref{eq:waveequation}) with an initial condition $\rho(r,0)$ and a no-slip boundary condition at $r=R$, is given by \cite{Duffy2015}
  \begin{align}
 \rho_1(r,t) = \frac{2}{R^2}\int_0^R\sum_{n=1}^\infty \rho_1(r',0)&\frac{J_0(k_n r / R) J_0(k_n r' / R)}{J_1(k_n)^2}\nonumber\\&\cos\left( \frac{ c_\text{k} k_n}{R}t\right) r' dr',
 \label{eq:analyticdensityprop}
 \end{align}
 where $k_n$ is the $n$th zero of $J_0(k) = 0$, $R$ is the channel height and $\rho_1(r',0)$ is the initial condition. 
 
 We initialized the fluid domain with a uniform density $\rho_0$ supplemented with a small disturbance $\rho_1$ field of width $w_d$ and amplitude $\rho_p$ in the center of the domain at $t_w=0$, where $t_w=\frac{c_k t}{w_d}$. In contrast to the oscillating droplet test, here we used the no-slip boundary condition at the top boundary, periodic boundary conditions along the sides and the mid-grid specular reflection boundary condition on the longitudinal $z$-axis. Furthermore, we set $\tau=0.6$ to be in the limit of small viscosity. Figure \ref{fig:propagatingdensitywave} shows the disturbance field $\rho_1$ as function of the position $r$ at different time stages for both the simulation and analytic (\ref{eq:analyticdensityprop}). The initial density disturbance causes two propagating waves in the system: one wave traveling towards the longitudinal axis and one traveling away from the longitudinal axis. Figure \ref{fig:propagatingdensitywave} clearly shows that mass is conserved in our simulation: there is a mass buildup in the wave traveling towards the longitudinal axis and mass loss for the wave traveling away from the longitudinal axis. We observe that the perturbed density field in the simulation is in excellent agreement with the analytical solution. There is a small departure from the analytical solution at $t_w=3.98$ as shown in figure \ref{fig:propagatingdensitywave}. This departure is most likely caused by the dissipative nature of the LBM with $\tau=0.6$, which is not described by (\ref{eq:waveequation}).
 \section{Conclusion and Discussion}
 We presented an axisymmetric LBM for high density ratio multiphase flows. The method is capable of achieving liquid-to-gas density ratios up to $10^3$ and higher. In order to recover the axisymmetric multiphase mass and momentum conservation equations appropriate source terms are introduced in the lattice Boltzmann evolution equation and in the multiphase force. The source terms in the evolution equation ensure the correct modeling for the mass, momentum and viscous tensors, while the source terms in the multiphase force ensure a correct surface tension in the model.
 
 We validated the model by comparing the Young-Laplace pressure of the axisymmetric simulation with the full 2D simulations. We observe a maximum relative error in the Laplace pressure of $3.2$\%. Then, we validated the dynamics of an axially symmetric oscillating droplet with a known analytical solution. Here, we observe a maximum relative error of $8$\%. Finally, we validate that a propagating density wave moving towards and away from the longitudinal z-axis correctly conserves mass and found excellent agreement with the inviscid analytical solution.

In our axisymmetric formulation of the LBM we used correction terms for the multiphase force that scale with the lattice speed of sound. We substitute the thermodynamic speed of sound for the lattice speed of sound in this correction term. A mathematical proof of the validity of this substitution cannot be provided at this moment, however we observe better consistency with the fully 2D counterpart. The ambiguous nature of the meaning of the speed of sound in this correction term is caused by imposing a non-ideal EOS. As a result, the original lattice Boltzmann formulation is violated where a constant speed of sound is assumed.
 
 \section{Acknowledgements}
  We acknowledge useful discussion with Matteo Lulli. This work is part of an Industrial Partnership Programme of the Foundation for Fundamental Research on Matter (FOM), which is financially supported by the Netherlands Organization for Scientific Research (NWO). This research programme is cofinanced by ASML.

\end{document}